\documentclass{PoS}

\title{Investigation of the $\mathbf{U_A(1)}$ in high temperature QCD on the lattice}

\ShortTitle{Investigation of the $U_A(1)$ in high temperature QCD on the lattice}

\author{\speaker{Sayantan Sharma}\\
        Fakult\"{a}t f\"{u}r Physik, Universit\"{a}t Bielefeld, D 33615, Germany\\
        E-mail: \email{sayantan@physik.uni-bielefeld.de}}

\author{Viktor Dick\\
        Fakult\"{a}t f\"{u}r Physik, Universit\"{a}t Bielefeld, D 33615, Germany\\
        E-mail: \email{viktor@physik.uni-bielefeld.de}
}
        
\author{Frithjof Karsch\\
         Fakult\"{a}t f\"{u}r Physik, Universit\"{a}t Bielefeld, D 33615, Germany~\\ \& Brookhaven National  Laboratory, Upton NY 11973, USA\\
        E-mail: \email{karsch@physik.uni-bielefeld.de}
}        
        
\author{Edwin Laermann\\
        Fakult\"{a}t f\"{u}r Physik, Universit\"{a}t Bielefeld, D 33615, Germany \\
        E-mail: \email{edwin@physik.uni-bielefeld.de}
} 
        
\author{Swagato Mukherjee\\
        Brookhaven National Laboratory, Upton NY 11973, USA\\
        E-mail: \email{swagato@bnl.gov}
}
\abstract{
In this project we study the effect of the $U_A(1)$ anomaly for $(2+1)$-flavour 
QCD at high temperature. 
We apply the overlap operator as a tool to probe the topological properties of 
gauge field configurations which have been generated within the Highly Improved 
Staggered Quark (HISQ) discretization scheme on lattices of size $32^3\times 8$ 
with $m_l/m_s=1/20$, commonly used for the study of QCD thermodynamics.
Although we have at present, only results for one value of the quark masses and 
thus cannot monitor the change of the eigenvalue distributions with the light quark mass,
the distribution of the low-lying eigenvalues of the overlap operator suggests that the 
$U_A(1)$ is not restored effectively even at 1.5 times the pseudo critical temperature. 
The corresponding low-lying eigenmodes show localization properties.
}

\FullConference{31st International Symposium on Lattice Field Theory LATTICE 2013\\
                 July 29 - August 3, 2013\\
                 Mainz, Germany}

\begin{document}

\section{Introduction}
The phase diagram of $N_f=2$ QCD at zero baryon density is still not clearly understood. Renormalization
group studies on sigma models with the same symmetries as QCD, suggest that the chiral phase 
transition for $N_f=2$ QCD can be continuous if the effective breaking of the $U_A(1)$ is 
sufficiently large, comparable to its zero temperature value~\cite{pw,bpv}. It is generally expected that 
the transition is first order if $U_A(1)$ is restored at the chiral transition temperature, although a second 
order transition may still be possible~\cite{bpv}. For $N_f=2+1$ flavours with physical quark masses, the masses of the light quarks $m_{u,d}<<m_s<\Lambda_{QCD}$ so the remnant chiral symmetry of two light quark 
flavours is still relevant near the chiral crossover. It is thus interesting to check the effects 
of the $U_A(1)$ breaking due to the two light flavours near the chiral crossover. 

Lattice gauge theory is the most effective non-perturbative tool that has helped in elucidating the 
phase diagram of QCD. It has been used in the recent years to understand the effects of $U_A(1)$ 
on the phase diagram of $N_f=2$ QCD. For two flavour QCD with dynamical overlap fermions, it was 
observed that $U_A(1)$ may be restored near the chiral transition~\cite{jl1}. A similar result was reported in this conference with the Optimal Domain Wall Fermions~\cite{twc}. These studies use relatively small lattices. However, with Highly Improved Staggered Quarks(HISQ) on larger and finer lattices, it was 
reported that $U_A(1)$ may be broken even at $1.5~T_c$, $T_c$ being the chiral cross-over temperature~\cite{ohno}.
Recent simulations with dynamical domain wall fermions with physical~\cite{dw1} as well as 
with slightly heavier quark masses, corresponding to $m_\pi=200$~MeV~\cite{dw2}, also suggest that 
$U_A(1)$ is broken in and above the chiral cross-over region. In view of the importance of this problem, evident from the activities followed by different groups, we tried to approach this problem in a 
different way. We use the configurations generated within the HISQ discretization scheme for fermions, 
for a large lattice volume and for almost physical quark masses and investigate 
their topological properties by means of the overlap Dirac operator, to understand the fate of $U_A(1)$ . These HISQ ensembles are compatible with $O(2)$ scaling for the chiral condensate near the cross-over region~\cite{hoteos} corresponding to the remnant chiral symmetry for the light quarks. This gives us a hint that these configurations could also be sensitive to the effects of $U_A(1)$ due to the light quarks.

\section{The Set-up}
Counting the topological objects in QCD configurations on a lattice is a difficult problem. 
In continuum QCD, if we can count the difference between the fermion zero modes with right and left chiralities, by the index theorem this difference is related to the topological charge. To extend this to the lattice, we need fermions which have exact chiral symmetry at finite lattice spacings. The overlap fermions~\cite{neunar} are the only known fermion discretization which have an exact chiral symmetry on the lattice. The lattice Dirac operator for the overlap fermions is given as, 
\begin{equation}
 D_{ov}=M\left[1+\gamma_5 sgn(\gamma_5 D_W(-M))\right]~,~sgn(\gamma_5 D_W(-M))=\gamma_5D_W(-M)/\sqrt{D_W^\dagger(-M)D_W(-M)}
\end{equation}
where $D_W$ is the Wilson-Dirac operator with a negative mass parameter $M\in [0,2)$. 

The overlap operator satisfies an index theorem~\cite{hln}. This ensures that by counting the number of zero modes of the overlap operator, we measure the topological charge of the underlying configurations.  In this work, we compute the eigenvalues and eigenvectors of the overlap fermion 
operator on the 2+1 flavour QCD configurations generated with the HISQ fermions. 
With the overlap operator we can not only identify the exact zero modes but distinguish these from the near zero modes. Also, the infrared region of the eigenvalue spectrum of the Dirac operator provides information about the chiral symmetry and the $U_A(1)$ breaking.
In terms of the eigenvalue density $\rho(\lambda,m)=T/V \langle \sum_k \delta(\lambda-\lambda_k(m))\rangle$, where $\lambda_k(m)$ are the eigenvalues of the Dirac operator with quark mass $m$,  the chiral condensate and the observable $\chi_\pi-\chi_\delta$ are given as,
\begin{equation}
 \label{eqn:ppbar}
 \langle\bar\psi\psi\rangle\overset{V\rightarrow \infty}{\rightarrow}\int_0^\infty d \lambda \frac{2m~\rho(\lambda,m)}{\lambda^2+m^2}~~~,~~~
\chi_\pi-\chi_\delta\overset{V\rightarrow \infty}{\rightarrow}\int_0^\infty d \lambda \frac{4m^2~\rho(\lambda,m)}{(\lambda^2+m^2)^2}~.
\end{equation}
In the limit of infinite volume and massless quarks, the chiral condensate 
depends on the density of eigenmodes near the origin. If ~$U_A(1)$ is restored simultaneously when 
the chiral symmetry is restored, then $\chi_\pi-\chi_\delta=0$. 
In the thermodynamic limit, one of the ways to realize this possibility is through opening of a gap in the 
infrared region of the eigenvalue spectrum. On the other hand, a possibility to have $U_A(1)$ broken
when chiral symmetry is restored is to have a finite density of near-zero modes with the 
eigenvalue distribution going as $\rho(\lambda)\sim m^2\delta(\lambda)$ as one approaches 
the chiral limit.

\section{Numerical details}
The set of $(2+1)$-flavour HISQ configurations used in this work were generated by the HotQCD
collaboration~\cite{hoteos}. The strange quark mass $m_s$ is set to the physical value and the light quark mass to $m_l=m_s/20$ which corresponds to a Goldstone pion mass of $160$~MeV.  The size of the lattice is $32^3\times 8$. We studied 3 sets of configurations at $1.04, 1.2$ and $1.5~T_c$, where $T_c$ is the pseudo-critical temperature for the chiral symmetry restoration and is about $159(2)$ MeV for $N_T=8$. We considered 100 configurations  for the two lowest temperatures and 161 for the highest temperature, each separated by 100 trajectories, and computed the eigenvalues of the overlap Dirac operator on them. 

For implementing the sign function in the overlap operator, we computed the lowest 20 eigenvectors of $D_W^\dagger D_W$ using the Kalkreuter-Simma(KS) Ritz algorithm~\cite{ks}. We computed the sign function for these low modes explicitly and for the higher modes, the sign function was approximated by a Zolotarev rational function. The number of terms in the Zolotarev function was kept to be 15. The overlap operator satisfies the Ginsparg-Wilson(GW) relation upto a precision of $10^{-10}$ and the square of the sign function deviated from identity by about  $10^{-9}$ on average. 

The 100 lowest eigenvalues at $1.04~T_c$ and 50 each at the other temperatures, of the overlap operator were computed using the KS algorithm with $D_{ov}^\dagger D_{ov}$. The convergence criterion for the KS algorithm was set to be $\epsilon^2<10^{-8}$. The corresponding fractional error on the zero eigenvalues
was $10^{-3}$ and for the others were better than $10^{-6}$. From partially-quenched 
studies, it is known that the choice of the parameter $M$ in the overlap operator is important because for 
certain choices of $M$, the corresponding $D_W^\dagger D_W$ could have very small eigenvalues leading 
to the presence of spurious zero modes in the overlap operator~\cite{ovm}. We verified that 
for configurations with 
no zero modes, the choice of the parameter $M$ did not affect the eigenvalues significantly 
within the precision with which these were determined. Moreover, for configurations with a finite number of zero modes of the overlap operator, we choose $M$ such that the sign function and the GW relation were determined with the highest accuracy, ensuring that we do not observe spurious eigenvalues. Except for a few cases, the 
parameter $M$ is chosen to be $1.8$. 
The implementation of the overlap operator and the computation of the eigenvalues were carried out on the 
GPU cluster at Bielefeld University.

\section{Results}
\begin{figure}
\begin{center}
\includegraphics[scale=0.5]{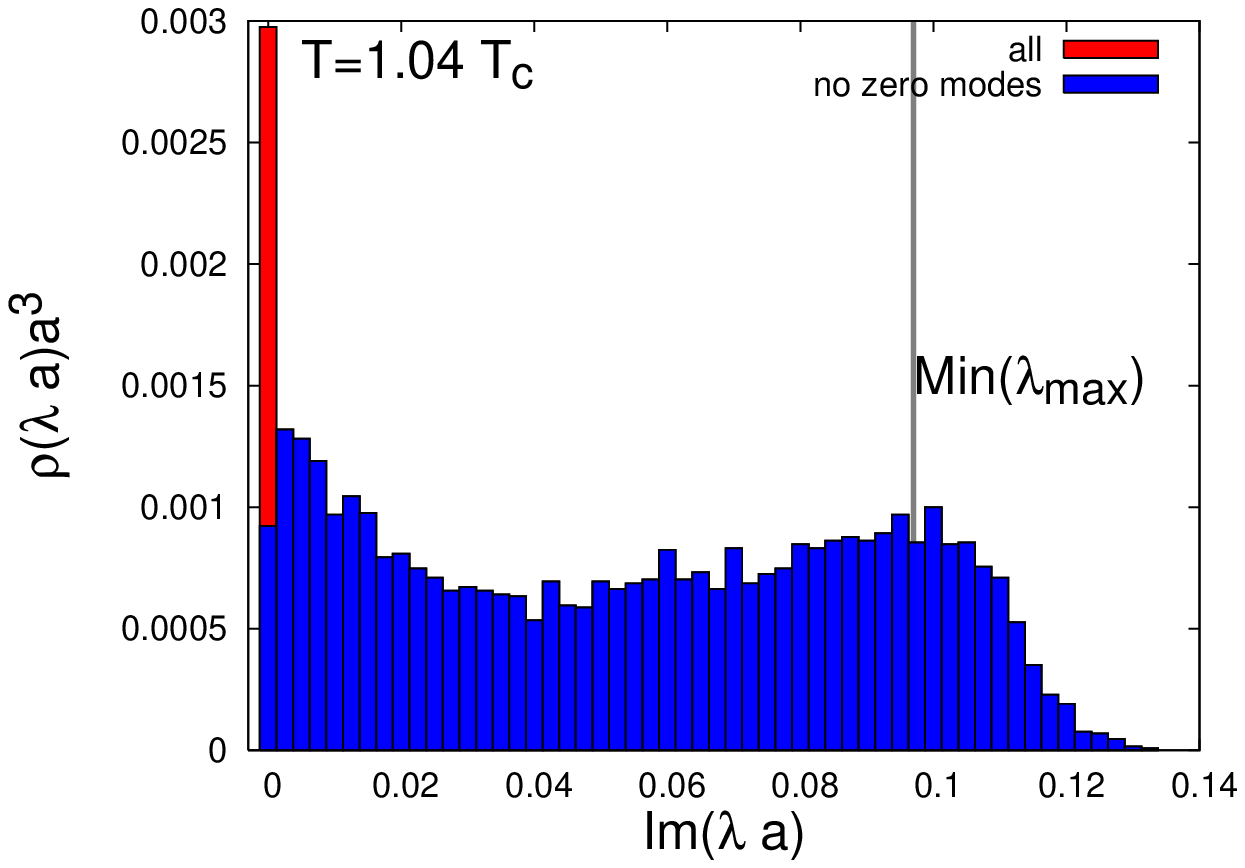}
\includegraphics[scale=0.5]{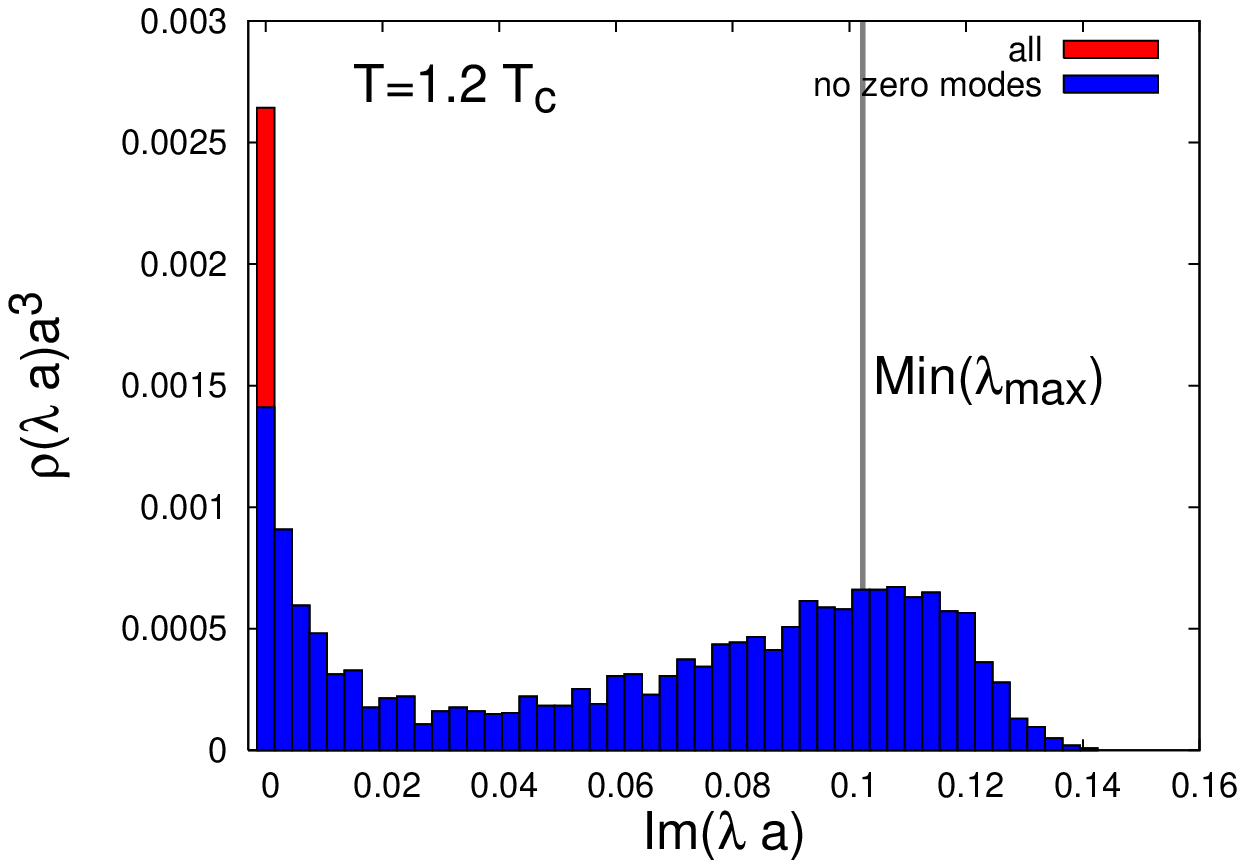}
\caption{The eigenvalue density of the overlap operator on HISQ sea configurations at temperatures 
1.04$T_c$(left panel) and $1.2T_c$ (right panel) as a function of imaginary part of the eigenvalue of the overlap operator. Min$(\lambda_{max})$ is obtained by taking  
the highest eigenvalue within each configuration and then taking the minimal value among these. Beyond this value, the eigenvalue density is no longer reliable.}
\label{eigval1}
\end{center}
\end{figure}

\begin{figure}
\begin{center}
\includegraphics[scale=0.5]{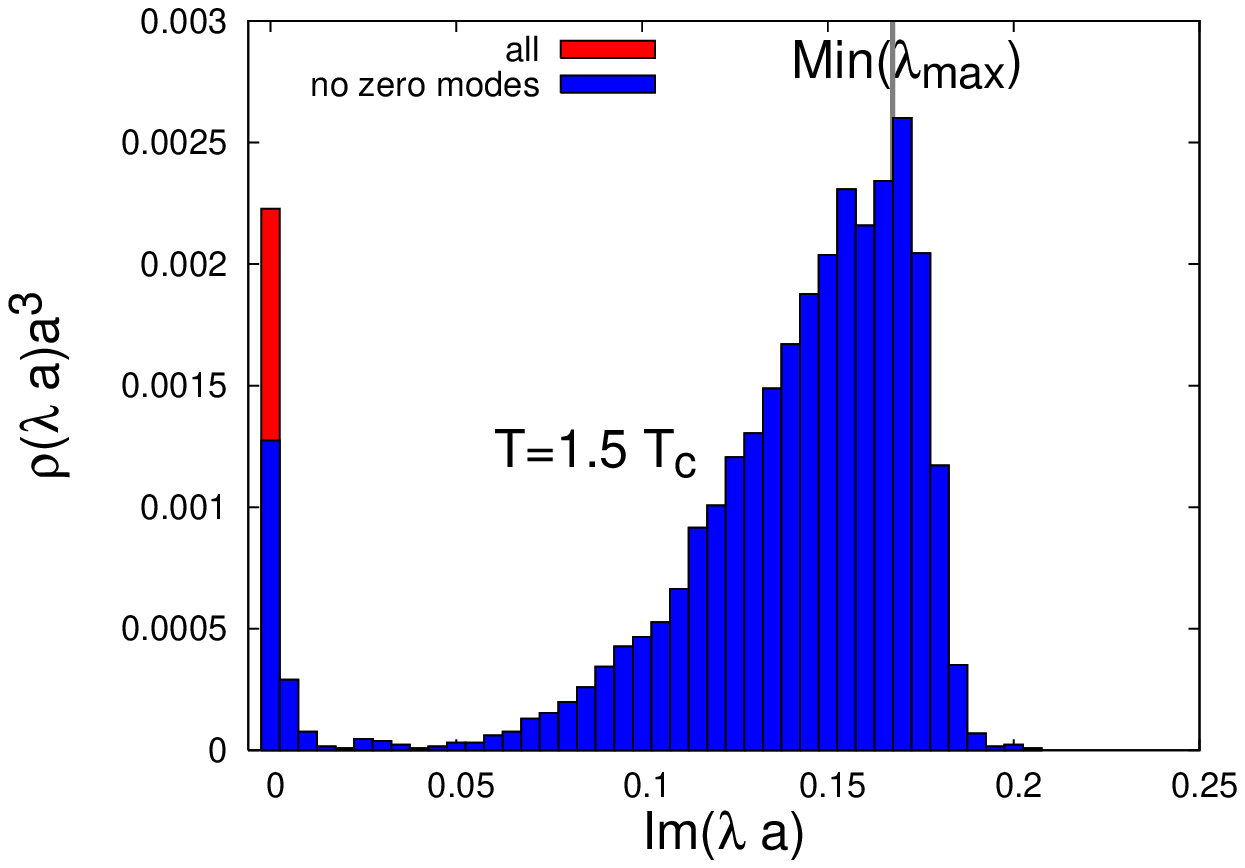}
\includegraphics[scale=0.5]{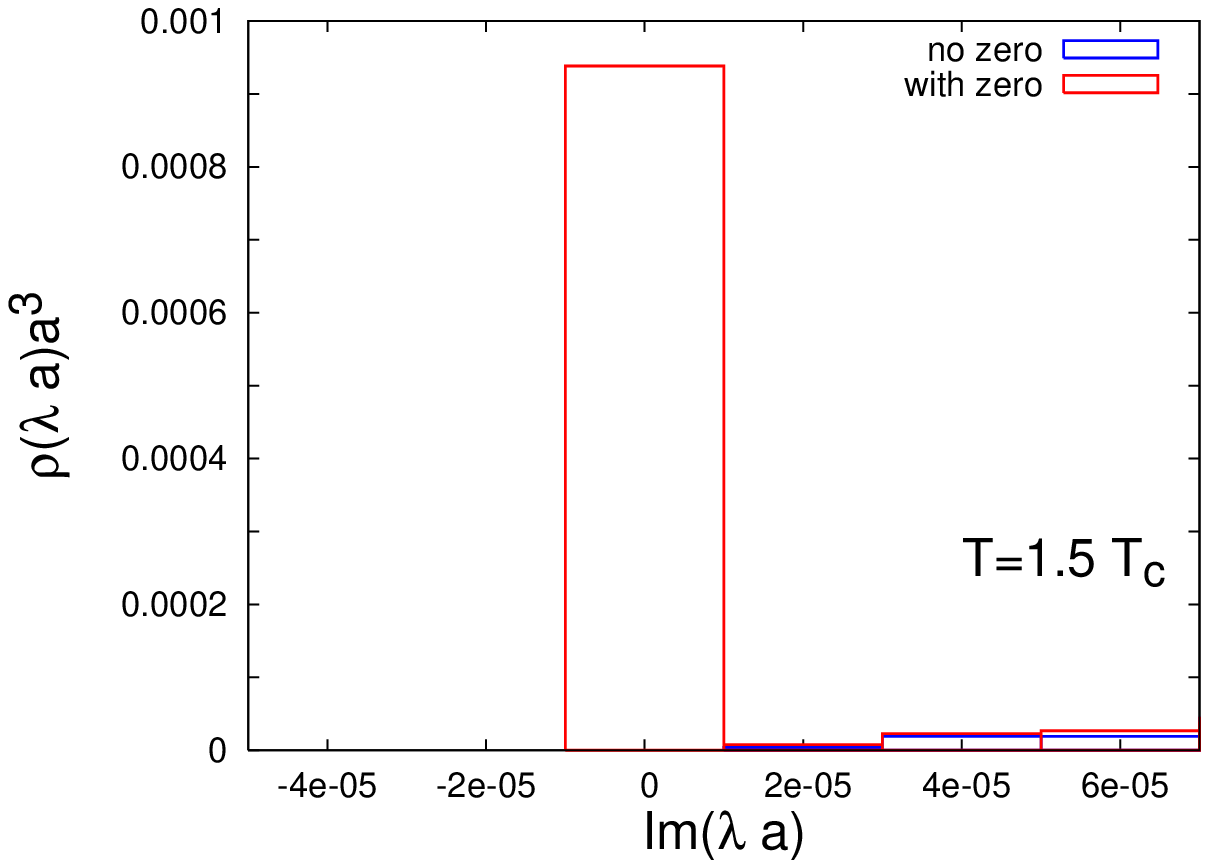}
\caption{The eigenvalue density of the overlap operator on HISQ sea configurations at 
1.5$T_c$(left panel) and the separation of zero and near-zero modes shown for the same temperature 
(right panel).}
\label{eigval2}
\end{center}
\end{figure}

We show our results for the eigenvalue distribution of the overlap operator on HISQ sea configurations 
for all the three temperatures in Figures (\ref{eigval1}) and (\ref{eigval2}). The red histograms represent the eigenvalue distribution including the zero modes while the blue ones count only the non-zero eigenvalues. We find that the density of zero modes decreases as the temperature of the QCD medium is increased. This is consistent with the expectation that the number of topological objects should reduce with increasing temperature. 
We checked the distribution of the different topological sectors as a function of temperature by 
counting the fraction of the number of configurations which have $Q$ zero modes. We compared them 
with the distribution of the same fraction, now measured with the operator $F\tilde{F}$ by employing 
HYP smearing~\cite{ohno} on these configurations. 
We find a good agreement between these two different estimates. Both these methods confirm the 
presence of a number of configurations having a large number of zero modes even at $1.5~T_c$.
\begin{figure}
\begin{center}
\includegraphics[scale=0.5]{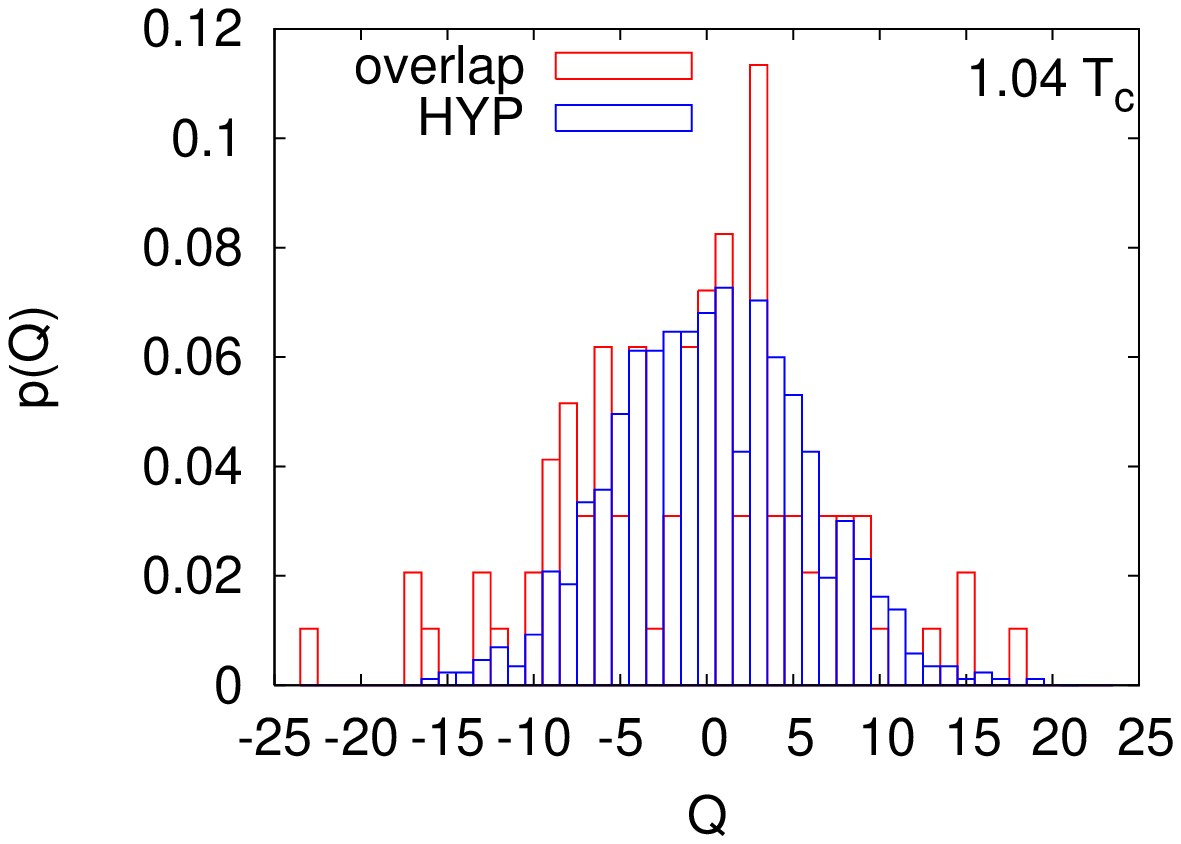}
\includegraphics[scale=0.5]{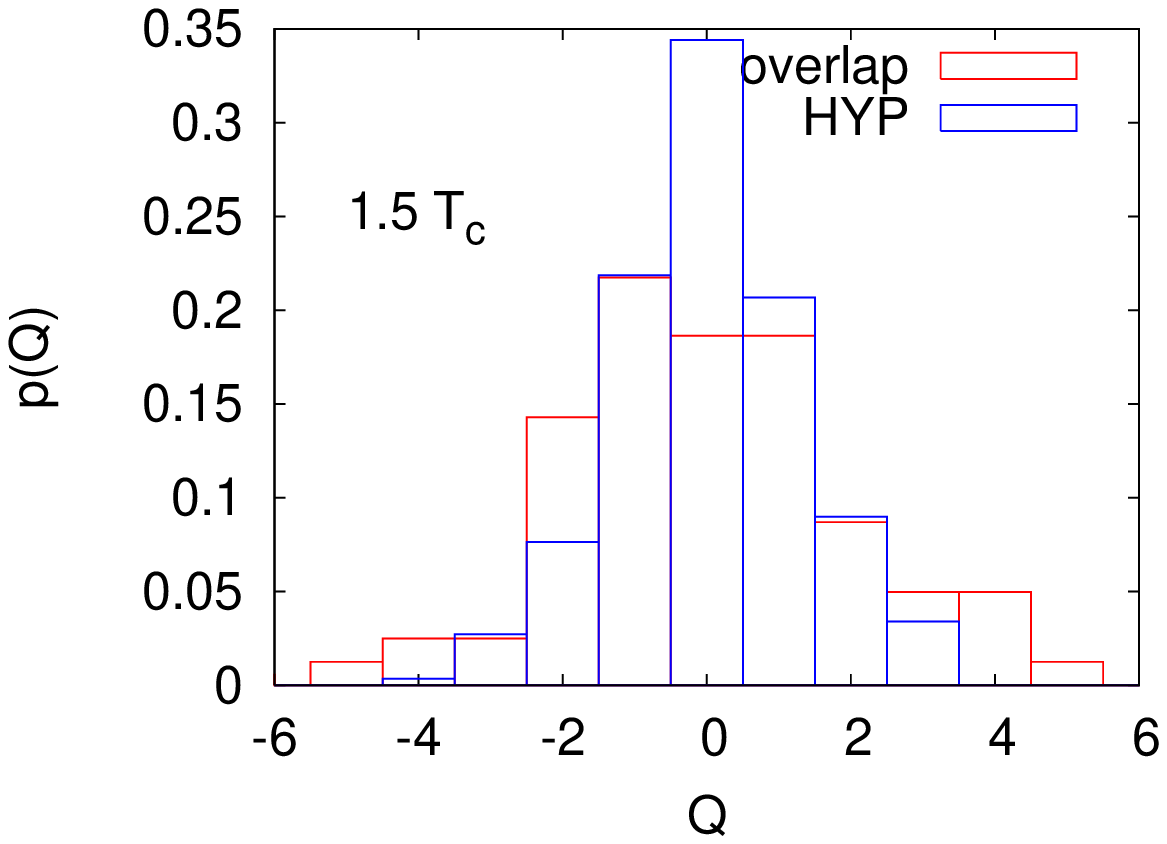}
\caption{The distribution of the topological charge obtained by counting the zero modes of the overlap operator 
and those by the operator $ F\tilde{F}$ after HYP smearing on the HISQ configurations at 1.04$T_c$(left panel) and $1.5~T_c$ (right panel). The number of configurations used with HYP smearing was $\mathcal{O}(10)$ larger.}
\label{hist}
\end{center}
\end{figure}

Using the overlap operator we can distinguish the zero modes from the near-zero modes very precisely as 
we show in the right panel of Figure (\ref{eigval2}). This is possible because we measure the eigenvectors 
$\psi_i$ of $D_{ov}^\dagger D_{ov}$. The zero modes always come with a definite chirality whereas the non-zero eigenvalues come in degenerate pairs with the corresponding chiralities being equal in magnitude but opposite in sign.  The density of these near zero modes too decreases with increasing temperature but it is non-vanishing
even at $1.5~T_c$, without any gap opening up in the eigenvalue spectrum. This gives us an 
evidence that $U_A(1)$ may not be restored even at $1.5~T_c$.

\begin{figure}
\begin{center}
\includegraphics[scale=0.5]{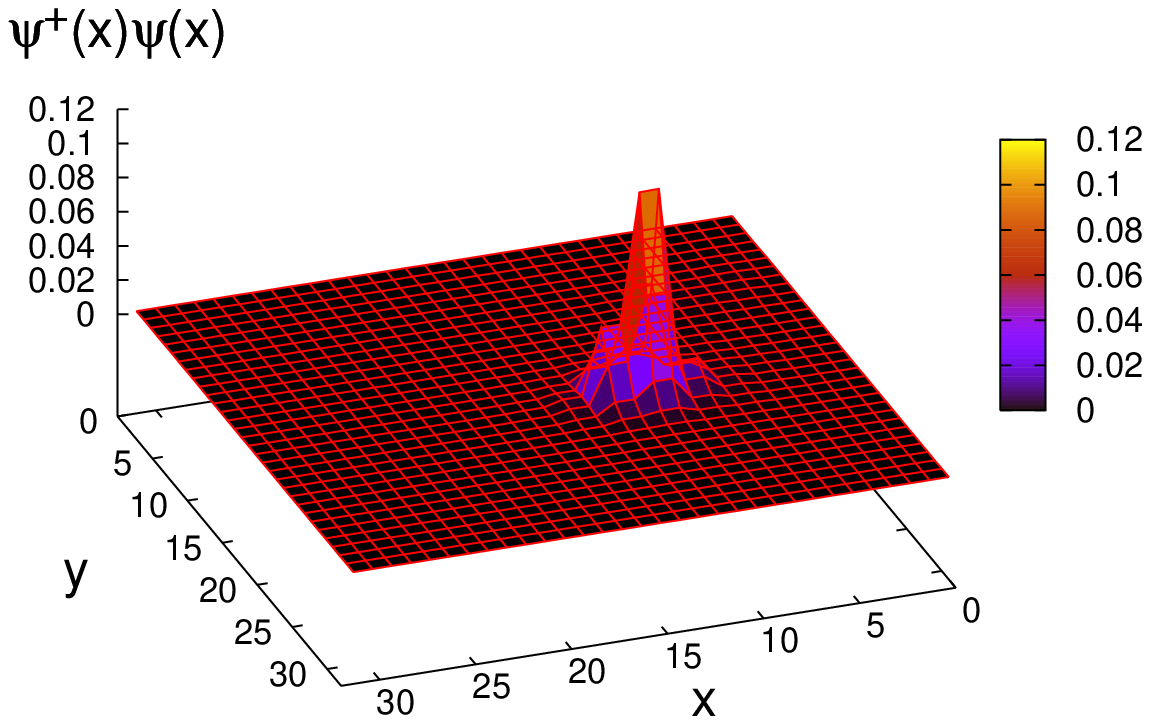}
\includegraphics[scale=0.5]{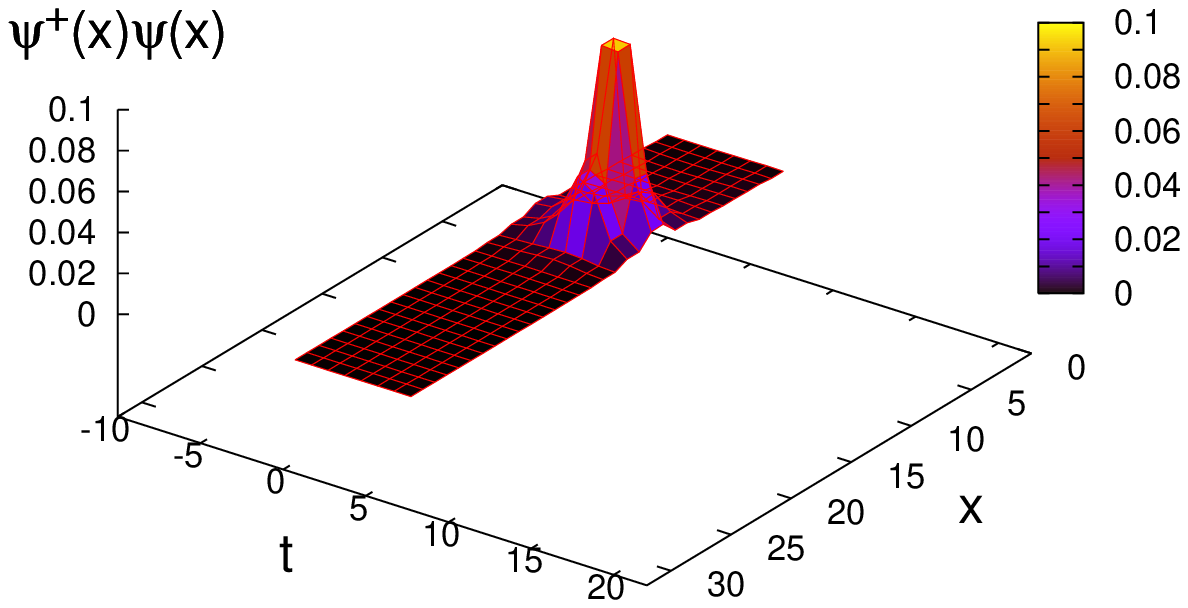}
\caption{The profile of a zero mode at $1.5~T_c$ along x-y plane(left panel) and x-t plane (right panel).}
\label{profile}
\end{center}
\end{figure}

In order to understand these findings in more detail, we first look at the properties of the 
zero modes at the highest temperature. We compute the probability density $\vert\psi_i\vert^2$ from the wavefunctions 
of the zero modes on different two dimensional coordinate slices by summing the wavefunction over the other two coordinates. The left panel of Figure (\ref{profile}) is the profile of a typical zero mode on the x,~y plane. The profile is localized along the spatial dimensions. The profile of the same zero mode along the x,~t directions, shown in Figure (\ref{profile}), indicates that the zero modes are localized along the temporal direction as well. The fermion wavefunctions are expected to be related to the non-trivial topological objects of the gauge fields, even at finite temperature. If we identify the profiles associated with the instantons, a rough estimate of the radius of these objects is about 0.2 fm in physical units. This is in agreement with the expectations from the dilute instanton gas model~\cite{shuryak}.

\begin{figure}
\begin{center}
\includegraphics[scale=0.5]{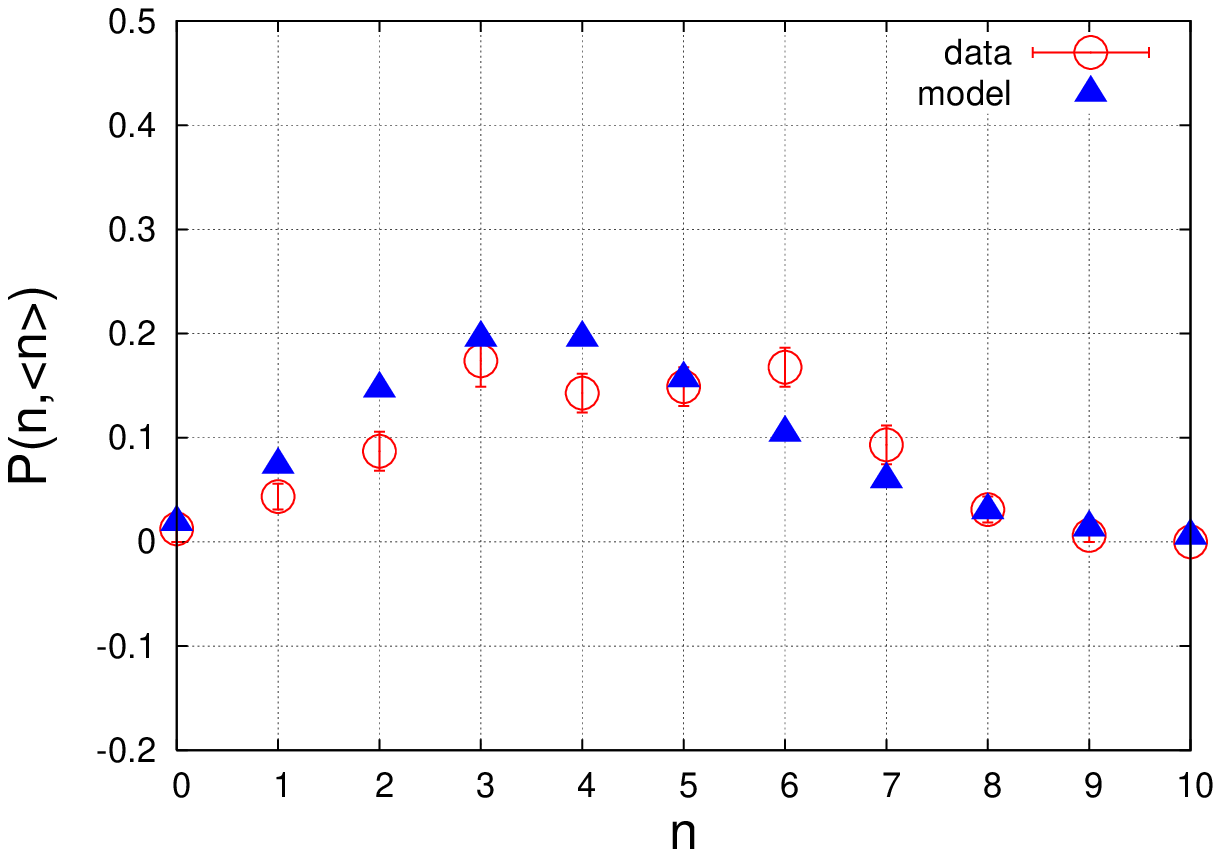}
\includegraphics[scale=0.5]{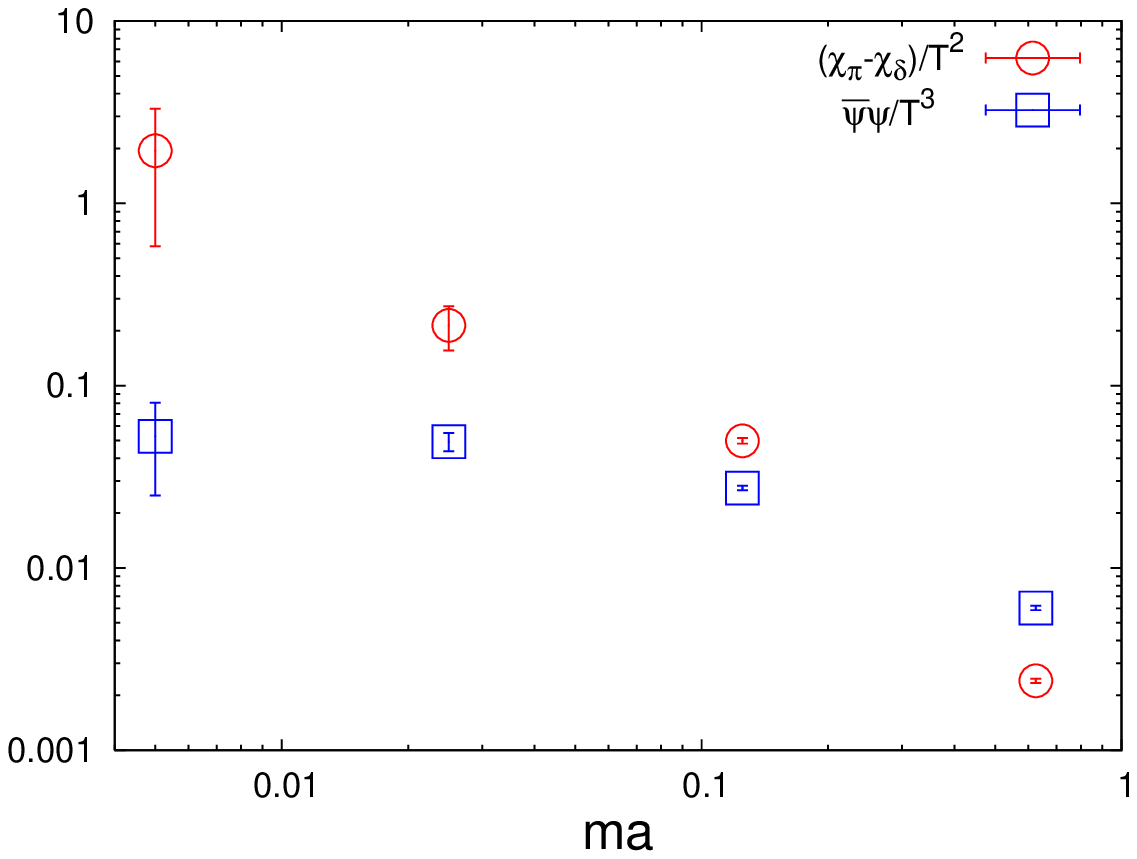}
\caption{The distribution of the net number of zero modes agrees well with the expectation with 
the Dilute Instanton Gas Model(left panel). In the right panel, we compare the values of the chiral condensate 
and $\chi_\pi-\chi_\delta$ for different values of the bare valence quark mass at $1.5~T_c$.}
\label{ndistr}
\end{center}
\end{figure}

The presence of near zero modes could be explained in this scenario of a dilute gas of instantons. The fermion wavefunctions localized with these instantons superimpose, resulting in the mixing of the lowest eigenvalues. This mechanism causes a spread of the zero eigenmodes in the infrared region of the eigenvalue spectrum. At $1.5~T_c$, we verify this assumption by counting all the eigenmodes with eigenvalues less than $Im(\lambda a) =0.049$, beyond which the contribution from the bulk eigenvalues dominate. The total number of these eigenfunctions per configuration is denoted by $n$. If these correspond to $n$ non-interacting instanton-anti-instantons, then the distribution of these objects should be Poisson like~\cite{ehn}, with 
the distribution function of the form, $P(n,\langle n\rangle)=e^{\langle n \rangle}\langle n\rangle/n!$. In order to verify this, we calculate the variance and the average value of the 
net number of topological objects and found that $\langle n\rangle=\langle n^2\rangle=4$.  
Moreover, we compare the distribution function $P(n,4)$ with 
the fraction of the total number of configurations having $n$ non-trivial topological objects. 
From Figure (\ref{ndistr}) it is evident that the topological objects indeed form a dilute gas of 
instantons. A similar observation was reported for pure SU(3) gauge theory~\cite{ehn} and with domain wall fermions~\cite{dw1}.

At $1.5~T_c$, the chiral symmetry is restored. This is also verified by computing 
the chiral condensate from the non-zero eigenvalues $\lambda_k^2$ of $D_{ov}^\dagger D_{ov}$ with different values of the valence quark mass,~$\hat m=ma$, given as 
$ \langle\bar\psi\psi\rangle=\sum_{|\lambda_k|>0}\frac{2 T}{V}\left\langle\frac{\hat m(4-\lambda_
k^2)}{\left[\lambda_k^2(1-\hat m^2)+4 \hat m^2\right]}\right\rangle$.
In the same way we also measure $\chi_\pi-\chi_\delta=\sum_{|\lambda_k|>0}\frac{4 T}{V}\left\langle\frac{\hat m^2(4-\lambda_k^2)
^2}{\left[\lambda_k^2(1-\hat m^2)+4 \hat m^2\right]^2}\right\rangle$ and found out that as we approach the chiral limit, $(\chi_\pi-\chi_\delta)/T^2$ is significantly larger than the chiral condensate supporting that the $U_A(1)$ is still broken in the chiral symmetry restored phase.

\section{Conclusions}
In this work, we studied the fate of the $U_A(1)$ in QCD on a large volume lattice. We use the overlap operator 
on HISQ sea configurations to identify unambiguously the number of zero modes present. We find a 
significant presence of zero and near-zero modes even at $1.5~T_c$, with no gap opening up in the 
infrared part of the eigenvalue spectrum indicating that $U_A(1)$ may not be restored. Though the 
contribution of the zero modes to the physical observables like the chiral condensate would vanish in the thermodynamic limit, the contribution of the near-zero modes would still survive. We check that these near zero modes do not contribute significantly to the chiral condensate at $1.5~T_c$ as expected, but the observable $\chi_\pi-\chi_\delta$ is finite and significantly larger than the chiral condensate for a wide range of valence quark mass. This further supports our claim that $U_A(1)$ may not be effectively restored 
even at $1.5~T_c$. We also study the possible mechanism of generation of such a large number of near-zero 
modes and our study suggests that the high temperature phase of QCD may well be described as a 
dilute gas of instantons. 

\section{Acknowledgements}
We acknowledge financial support from the GSI BILAER grant and from the European Union under 
Grant Agreement number 238353.

\end{document}